\documentclass{webofc}
\usepackage[varg]{txfonts}   
\newcommand{\bmodes}{{\it B}-modes}
\newcommand{\emodes}{{\it E}-modes}
\begin{document}
\title{Polarization angle accuracy for future CMB experiments}
\subtitle{The COSMOCal project and its prototype in the 1 mm band}
%
%
   \author{\lastname{A. Ritacco}\inst{1,2}\thanks{e-mail: alessia.ritacco@inaf.it} \and
        \lastname{L. Bizzarri}\inst{2,3} \and
        \lastname{F. Boulanger}\inst{2} \and
        \lastname{M. P\'{e}rault}\inst{2} \and
        \lastname{J. Aumont}\inst{13} \and
        \lastname{F. Bouchet}\inst{9} \and
        \lastname{M. Calvo}\inst{11} \and
        \lastname{A. Catalano}\inst{6} \and
        \lastname{D. Darson}\inst{2} \and
        \lastname{F.X. D\'{e}sert}\inst{10} \and
        \lastname{J. Errard}\inst{12} \and
        \lastname{A. Feret}\inst{5} \and
        \lastname{J. F. Mac\'{i}as-P\'{e}rez}\inst{6} \and
        \lastname{B. Maffei}\inst{4} \and
        \lastname{A. Monfardini}\inst{11} \and
        \lastname{L. Montier}\inst{13} \and
        \lastname{M. Murgia}\inst{1} \and
        \lastname{P. Morfin}\inst{2}\and
        \lastname{F. Nati}\inst{3}\and
        \lastname{G. Pisano}\inst{7}\and
        \lastname{N. Ponthieu}\inst{10}\and
        \lastname{J. L. Puget}\inst{2}\and
        \lastname{S. Savorgnano}\inst{6}\and
        \lastname{B. Segret}\inst{7}\and
        \lastname{K. Schuster}\inst{14}\and
        \lastname{J. Treuttel}\inst{5}\and
        \lastname{M. Zannoni}\inst{3}
          }
\institute{%
INAF-Osservatorio Astronomico di Cagliari, Via della Scienza 5, 09047 Selargius, IT
\and
Laboratoire de Physique de l’$\acute{\rm E}$cole Normale Sup$\acute{\rm e}$rieure, ENS, Universit$\acute{\rm e}$ PSL, CNRS, Sorbonne Universit$\acute{\rm e}$, Universit$\acute{\rm e}$ de Paris, 75005 Paris, France
\and 
Department of Physics, Universit\'{e} di Milano-Bicocca, Piazza della Scienza, 3, 20126, Milan, Italy
\and
IAS, CNRS, Universit\'{e} Paris-Saclay, CNRS, B\^{a}t. 121, 91405 Orsay, France
\and
LERMA, Observatoire de Paris-PSL, 61 Avenue de l’Observatoire 75014
Paris France
\and
Univ. Grenoble Alpes, CNRS, LPSC-IN2P3, 53, avenue des Martyrs, 38000 Grenoble, France
\and
CENSUS, Observatoire de Paris-PSL, Universit\'{e} PSL, 92195 Meudon, France
\and
Dipartimento di Fisica, Sapienza Universit\`{a} di Roma, 00185 Roma, Italy
\and
Institut d’Astrophysique de Paris, CNRS \& Sorbonne Universit\'{e} (UMR7095), 75014 Paris, France
\and
Univ. Grenoble Alpes, CNRS, IPAG, 38000 Grenoble, France
\and
Universit\'{e} Grenoble Alpes, CNRS, Institut N\'{e}el, France
\and
Universit\'{e} Paris Cité, CNRS, Astroparticule et Cosmologie, F-75013 Paris, France
\and
Universit\'{e} de Toulouse, UPS-OMP, CNRS, IRAP, 31028 Toulouse, France
\and
Institut de Radioastronomie Millim\'{e}trique (IRAM), 38406 Saint Martin d’H\`{e}res, France
}

\abstract{%
The Cosmic Microwave Background (CMB) radiation offers a unique window into the early Universe, facilitating precise examinations of fundamental cosmological theories. However, the quest for detecting \bmodes~in the CMB, predicted by theoretical models of inflation, faces substantial challenges in terms of calibration and foreground modeling. The COSMOCal (\textbf{CO}smic \textbf{S}urvey of \textbf{M}illimeter wavelengths \textbf{O}bjects for CMB experiments \textbf{Cal}ibration) project aims at enhancing the accuracy of the absolute calibration of the polarization angle $\psi$ of current and future CMB experiments.
The concept includes the build of a very well known artificial source emitting in the frequency range [20-350] GHz that would act as an absolute calibrator for several polarization facilities on Earth. A feasibility study to place the artificial source in geostationary orbit, in the far field for all the telescopes on Earth, is ongoing. In the meanwhile ongoing hardware work is dedicated to build a prototype to test the technology, the precision and the stability of the polarization recovering in the 1 mm band (220-300 GHz). High-resolution experiments as the NIKA2 camera at the IRAM 30m telescope will be deployed for such use. Once carefully calibrated ($\Delta$ $\psi$ $<$ 0.1$^{\circ}$) it will be used to observe astrophysical sources such as the Crab nebula, which is the best candidate in the sky for the absolute calibration of CMB experiments. 
}
\maketitle
\section{Introduction}
\label{intro}
The Cosmic Microwave Background (CMB) radiation offers unique insights into the structure and evolution of the early Universe. Recent precision measurements have confirmed the standard cosmological model and supported the concept of cosmic inflation \citep{planckI2018}.

A significant goal in CMB research today is the detection of the primordial gravitational waves imprint on the CMB polarization, as curl-like patterns known as \bmodes~\citep{HU1997323} and predicted by the inflation theory. The CMB polarization also exhibits a coherent and symmetrical signal originating mainly from density fluctuations in the primordial Universe, referred to as the \emodes. Detecting the CMB \bmodes~signal is challenging due to their very weak amplitude, which is more than an order of magnitude weaker than the \emodes, and as so the high sensitivity demanded at the instrumental level. Furthermore, at the data analysis level, two critical aspects (among others) need to be addressed: i) the miscalibration of the polarization angle, which can induce leakage from $E$- to \bmodes~\cite{aumont2020}; ii) the presence of galactic foreground due to dust and synchrotron emission that have not yet been properly modeled \cite{ritacco2023, nico2018}.
The requirement is set on the level of the primordial \bmodes~amplitude, which is parameterized by the tensor-to-scalar ratio parameter $r$, and it is expected to be very weak, of the order of 10$^{-3}$ by the most accredited inflationary models to date \cite{litebird2023}. An accurate and unbiased determination of $r$ will tell us how much gravitational waves (tensor perturbations) there are compared to density fluctuations (scalar perturbations) in the early Universe. This would be a confirmation of the Inflation theory and would open the path to new discoveries in the physics of the primordial Universe.

In this article, we summarize recent results that challenge the unbiased detection of CMB \bmodes, and we introduce a prototype development of a calibration system, as part of a more extensive project named COSMOCal, to be tested at the IRAM 30m telescope with the NIKA2 camera \cite{nika2}. The final goal of this project study is to demonstrate that the chosen technology and calibration strategy can ensure an accuracy in the absolute polarization angle reference of $\Delta\psi$ < 0.1 degrees, meeting the requirements for the next generation of CMB experiments.

\section{Scientific context}
\label{sec-2}
\subsection{Challenges derived from galactic dust emission}
\label{sec-2.1}
The CMB polarization is dominated by \emodes~associated with density fluctuations in the early universe.
An error on the calibration of polarization angles induces a leakage from \textit{E} to \bmodes~\cite{abitbol2016}.
The calibration of angles needs to be accurate enough to ensure that this leakage is smaller than the amplitude of primordial \bmodes.
 An angle error creates \bmodes~signal correlated on the sky with the dominant
\emodes. The baseline of LiteBIRD, as well as ground-based experiments, is to use the \textit{EB}
cross-spectrum to calibrate polarization angles \cite{litebird2023}. This approach is supported by the
fact that the \textit{EB} correlation is null for the standard cosmological model. However, alternative cosmological
models of great scientific interest predict that the Universe is birefringent, which would make the
cosmological \textit{EB} non-zero \cite{minami_komatsu}. This phenomenon indeed naturally transforms $E$ modes into \bmodes. This transformation arises from the parity-violating extensions of the standard model, as proposed by \cite{carroll1990} and \cite{pospelov2009}.

Furthermore, galactic foregrounds can also create a non-zero \textit{EB} signal.
A recent refined multi-frequency analysis highlighted that current foreground component separation models are not sufficient to explain the spatial variation of dust properties in polarization \cite{ritacco2023}. The authors present a cross power spectra analysis performed on full-sky spatial residuals extrapolated from \textit{Planck} observations. This is used to quantify the correlations \textit{EE} and \textit{BB} between two different data sets.
These residual maps are obtained by subtracting an averaged full-sky dust SED (Spectral Energy Distribution).
To interpret the results obtained in polarization \textit{Planck} HFI data, the authors have used total intensity dust emission models to extract synthetic maps in polarization. A cross power spectra analysis has been performed to evaluate the correlation between real \textit{Planck} HFI maps and the synthetic maps extracted from a dust model. The results show that current dust models are only able to capture the structure of the interstellar medium in the galactic plane. At larger angular scales the correlation between polarization data and current dust models based is significantly lower than the correlation between two different polarization data sets.
This result highlights that, in polarization, current dust models are missing important information about the structure of the interstellar medium. 

In addition \textit{Ritacco et al. 2023} \cite{ritacco2023} shows that a crucial role in shaping the polarization dust SED is instead played by the variation of the polarization angle with frequency along the line of sight. 
A theoretical description has been proposed by \textit{Tassis et al. 2015, Vacher et al 2023} \citep{tassis2015, vacher2023b} and observed in discrete sets of sightlines by \textit{Pelgrims et al. 2021} \citep{pelgrims2021}. \textit{Ritacco et al. 2023} \cite{ritacco2023} found the extrapolation of this effect at the power spectrum level. 
In this case, the strong contribution to the total signal amplitude in polarization data can be attributed primarily to the correlation extrapolated from the residual maps of polarization angles. This provides a clear indication of the the need to consider polarization angles to model the dust SED in polarization. 


\subsection{Absolute calibration of the polarization angle}
The absolute calibration of the polarization angle is a significant challenge confronting upcoming ground-based, balloon-borne, and satellite CMB polarization experiments.
An alternative to the \textit{EB}-nulling strategy mentioned above involves using sky calibration through a well-established reference source: the Crab nebula (or Tau A), renowned as the brightest polarized astrophysical source in the microwave sky, particularly at angular scales of a few arc-minutes. The Crab nebula has undergone extensive investigations for this purpose in the past, with studies conducted by \cite{aumont2010} and \cite{macias2010}, and more recent by \cite{ritacco2018} and \cite{aumont2020}.
These latter studies estimated the weighted average of the polarization angle in a large range of frequencies [23-353] GHz, obtaining a statistical uncertainty of 0.27$^{\circ}$. However, this estimate is limited by the instrumental error, which is of order of  1$^{\circ}$ and implies the strong assumption that the polarization angle of the Crab nebula is constant in the wide range of frequencies considered. 
Clearly, nowadays in the era of precision cosmology, this level of uncertainty is insufficient. Enhancing this precision level by a factor of at least ten represents a challenging endeavor.

So, to interpret results from future CMB polarization experiments, independently from cosmological and
foregounds models, and instrumental systematics we need to develop a calibration approach independent of the observation of the
diffuse microwave emission. This effort is underway for ground-based experiments using reference sources
on the ground \cite{cornelison2022} or on drones \cite{nati2017} but these avenues do not apply to space experiments, in particular LiteBIRD. The COSMOCal project envisages to establish a common reference enabling cross-correlation between different polarization experiments, from Earth and space. The following section will present the idea and the current status of the project.
\section{COSMOCal project}
\label{sec-3}
The COSMOCal project aims at establishing a model independent method to cross-calibrate different polarization instruments at the precision required by current under development CMB experiments. 
The main idea behind the project is to develop a space mission providing a reference polarized signal over the 20-350 GHz
frequency range with a polarization angle known with an absolute precision of <0.1$^{\circ}$ with respect to sky coordinates. The central concept underpinning this endeavor involves placing the microwave source in space to be in the far-field of large telescopes and on a geo-stationary orbit (GEO) due to tracking limitations. First, the source will be used to calibrate polarization properties of a few ground-based
telescopes in Europe and Chile selected to cover the full frequency range. Second, these telescopes will be used to map the polarized emission of a set of astrophysical sources, comprising supernova remnants and radio galaxies. These sources will be chosen for their suitability in terms of sky coverage, as well as stability in both frequency and time dimensions. Last, the space mission will provide the CMB community and astrophysicists with a common reference tying the calibration of space and ground-based polarization observations at microwave frequencies. In addition to the feasibility study for the space mission, which has the advantage of ensuring stability and repeatability of the calibration, we are envisaging to test the COSMOCal prototype already built, extending its capabilities to a wider range of frequencies, using small aperture telescopes on the ground like the QUBIC experiment installed in Argentina or the SATs of the Simons Observatory in Chile. The cross calibration of those telescopes with the large aperture ones will immediately give us a hint on the feasibility of the long term project and will already allow us to advance on our goal to achieve the required accuracy on bright reference sources as the Crab nebula. This is part of the ongoing work and feasibility study. 
This project study encompasses several phases, ranging from initial technology assessment to the creation of a data analysis pipeline for cross-referencing observations between high-resolution ground-based telescopes and low-resolution Cosmic Microwave Background (CMB) telescopes, to be able to extend the absolute calibration to the LiteBIRD satellite in the future.
\subsection{Development of a prototype}
\begin{figure}[h]
\centering
\includegraphics[scale=0.5]{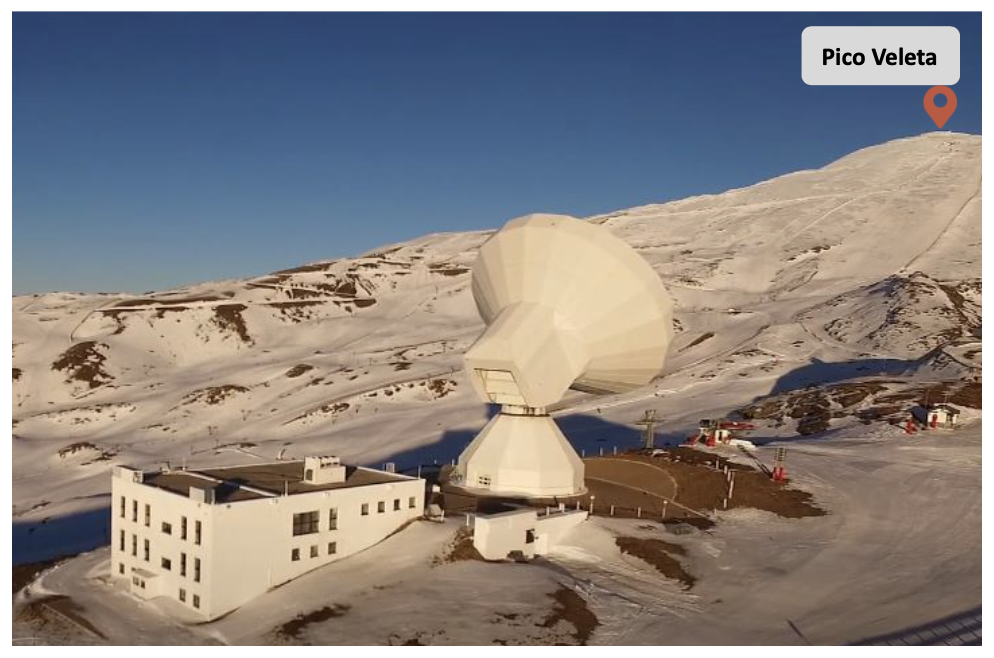}
\caption{Picture showing the IRAM 30m telescope and the position on the peak of the Pico Veleta mountain where the COSMOCal prototype will be placed.}
\label{fig-4}       
\end{figure}

\begin{figure}[h]
\centering
\includegraphics[scale=0.5]{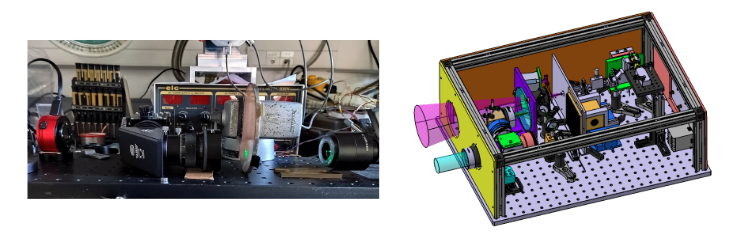}
\caption{\textit{Left}: picture taken during optical system's measurements at LPENS in Paris. \textit{Right}: mechanical drawing of the COSMOCal prototype.}
\label{fig-5}       
\end{figure}

Currently, the COSMOCal collaboration effort is focused in constructing a prototype designed to operate within the 220-300 GHz frequency band, which is
one of the most difficult frequency bands to target from Earth. This prototype is scheduled for testing at the LPSC laboratory in Grenoble during the upcoming months, and with the IRAM 30m telescope in Spain by 2024 using the NIKA2 camera. Fig.~\ref{fig-4} shows the IRAM 30m telescope site and where the artificial source is intended to be placed.
Achieving the required objective in calibration accuracy involves identifying and correcting all instrumental effects that affect the observations accuracy. 
The mechanical drawing of the COSMOCal prototype is shown on the right panel of Fig.~\ref{fig-5}. It consists of: a) a millimeter source chain with a resulting signal emitted at 265 GHz, which matches the central frequency of the 1mm band of NIKA2; b) an optical system that ensures the instantaneous reconstruction of the polarization orientation of the emitted mm signal.
The optical system includes: a laser illuminating a wire-grid polarizer, a flip mirror and a CCD camera. 
The polariser is used to ensure the purity of the polarization signal and the laser is used to image the orientation of its wires. The passage of the optical light will produce a diffraction pattern that is imaged by the CCD camera. The flip mirror is used to alternatively image ground references, as observed by the refractor telescope, placed around the IRAM 30m telescope. These references will allow us to establish the box’s orientation w.r.t the IRAM reference frame. 

Laboratory characterization on the millimeter source and the different optical components are ongoing, as pictured by the image shown on the left panel of Fig.~\ref{fig-5}. A more extensive description of the first results obtained will be discussed in an upcoming article (\textit{Ritacco, A. and COSMOCal coll. to be submitted} \cite{ritacco_cosmocal}).

\section{Conclusion}
This article provides a synthetic overview of some of the technical challenges associated with detecting Cosmic Microwave Background (CMB) radiation's \bmodes. One of these challenges pertains to the absolute calibration of the polarization angle. Achieving the level of accuracy required to prevent instrumental systematic effects from converting \emodes~ into \bmodes~ is quite challenging. Here, we propose a model-independent method for calibrating the absolute polarization orientation in CMB experiments, which may enable the exploration of phenomena such as Cosmic Birefringence, a topic that is currently highly debated (see \cite{diego-palazuelos}).

To address these challenges, the article discusses the development of a calibration system as part of the COSMOCal project. This project aims to attain an absolute polarization angle reference accuracy of less than 0.1 degrees, meeting the requirements for the next generation of CMB experiments. The article introduces the development of a prototype for testing at the IRAM 30m telescope with the NIKA2 camera.

The prototype's optical and electronic components are currently undergoing testing in the laboratories of the Paris Observatory, LPENS in Paris, and the IAS institute in Orsay. The assembled prototype will be tested at the LPSC laboratory in Grenoble next winter, using a similar millimeter camera as NIKA2. The measurements at the IRAM 30m telescope are expected to be conducted by 2024.
\section*{Acknowledgements} 
We acknowledge financial support from CENSUS, Observatoire de Paris-PSL. A.R. acknowledges financial support from the Italian Ministry of University and Research - Project Proposal CIR01\_00010. 

%
%
%

\end{document}